\documentclass[twocolumn,prb]{revtex4}
\usepackage{amsfonts}
\usepackage[T1]{fontenc}
\usepackage{amsmath,amsbsy,amssymb,graphicx}
\usepackage{times}

\begin{document}

\title{Merging of momentum-space monopoles by controlling magnetic field:\\
From cubic-Dirac to  triple-Weyl fermion systems}
\author{Motohiko Ezawa}
\affiliation{Department of Applied Physics, University of Tokyo, Hongo 7-3-1, 113-8656,
Japan}

\begin{abstract}
We analyze a generalized Dirac system, where the dispersion along the $k_{x}$
and $k_{y}$ axes is $N$-th power and linear along the $k_{z}$ axis. When we
apply magnetic field, there emerge $N$ monopole-antimonopole pairs beyond a
certain critical field in general. As the direction of the magnetic field is
rotated toward the $z$ axis, monopoles move to the north pole while
antimonopoles move to the south pole. When the magnetic field becomes
parallel to the $z$ axis, they merge into one monopole or one antimonopole
whose monopole charge is $\pm N$. The resultant system is a multiple-Weyl
semimetal. Characteristic properties of such a system are that the anomalous
Hall effect and the chiral anomaly are enhanced by $N$\ times and that $N$
Fermi arcs appear. These phenomena will be observed experimentally in the
cubic-Dirac and  triple-Weyl fermion systems ($N=3$).
\end{abstract}

\maketitle

\textit{Introduction: }Topological objects in the momentum space play
intriguing roles in the field of condensed matter physics\cite%
{Hasan,Qi,Hosur,JPSJReview}. Examples are monopoles, skyrmions and merons.
They have fascinating properties that are not shared by the corresponding
ones in the real space, since they are purely static because of the absence
of the kinetic energy. For instance, a monopole carrying a large magnetic
charge cannot exist in the real space due to a large Coulomb repulsion but
can in the momentum space. It is an interesting problem if we may generate $%
N $ monopole-antimonopole pairs and successively make them merged into one
monopole-antimonopole pair each of which carries monopole charge $\pm N$
just by controlling an external parameter.

Weyl semimetal is characterized by the monopole charge in the momentum space%
\cite{Murakami}. Characteristic features of Weyl semimetals are the
emergence of the anomalous Hall conductance\cite{Burkov,KK,Hosur}, the
chiral anomaly\cite{ABJ,Zyuzin,Huang,PYe,CLZhang} and the Fermi arc\cite%
{XWan,Hosur,SYXu}. Double- and triple-Weyl semimetals are generalization of
Weyl semimetals, where the monopole charges are $2$ and $3$, respectively%
\cite{CFang,BJYang,Xli,Huang2,Jian,Ahn}. It has theoretically been proposed
that quadratic- and cubic-Dirac insulators are also possible based on the
symmetry considerations\cite{BJYang}. The dispersion is parabolic or cubic
along the $k_{x}$ and $k_{y}$ directions, while it is linear along the $%
k_{z} $ direction. Very recently, it is shown that the cubic Dirac
semimetals would be materialized in quasi-one-dimensional transition-metal
monochalcogenides by first-principles calculations\cite{QLiu}.

In this paper we propose a simple model realizing a merging process of
monopoles in the momentum space. It is a generalized Dirac system, where the
dispersion along the $k_{x}$ and $k_{y}$ axes is $N$-th power and is linear
along the $k_{z}$ axis. Let us apply magnetic field. Unless its direction is
not along the $z$ axis, there emerge $N$ pairs of Weyl and anti-Weyl
fermions beyond a certain critical value of the field. As the direction
approaches the $z$ axis, these Weyl fermions with the positive monopole
charge move to the north pole, while those with the negative monopole charge
move to the south pole. When the magnetic field becomes parallel to the $z$
axis, they merge into one multiple-Weyl point whose monopole charge is $N$.
We show in such a system that the anomalous Hall effect and the chiral
anomaly are enhanced by $N$\ times, and furthermore that $N$ Fermi arcs
emerge. These phenomena will be observed experimentally in the quadratic and
cubic Dirac fermion systems.

\textit{Model Hamiltonian:} We investigate the following Hamiltonian in the
momentum space $(k_{x},k_{y},k_{z})$, 
\begin{equation}
H=\tau _{z}\left[ ta^{N}\left( k_{+}^{N}\sigma _{-}+k_{-}^{N}\sigma
_{+}\right) +t_{z}a_{z}k_{z}\sigma _{z}\right] +m\tau _{x}+\mathbf{h}\cdot 
\mathbf{\sigma },  \label{BasicHamil}
\end{equation}%
where $N$ is an arbitrary natural number; $k_{\pm }=k_{x}\pm ik_{y}$; $%
t,t_{z}$ are constants of energy dimension; $a,a_{z}$ are constants of
length dimension; $m$ is the mass parameter taken in energy dimension ($m>0$%
); $\mathbf{h}\cdot \mathbf{\sigma }$ represents the Zeeman energy due to
external magnetic field (let us call $\mathbf{h}$ magnetic field); $\mathbf{%
\sigma }$ and $\mathbf{\tau }$ are the Pauli matrices describing the spin
and pseudospin degrees of freedom, respectively; $\sigma _{\pm }=\sigma
_{x}\pm i\sigma _{y}$. For simplicity we set $t=t_{z}=1$ in what follows.
They can be easily recovered since they appear always in pairs with $a$ and $%
a_{z}$.

\begin{figure*}[t]
\centerline{\includegraphics[width=0.99\textwidth]{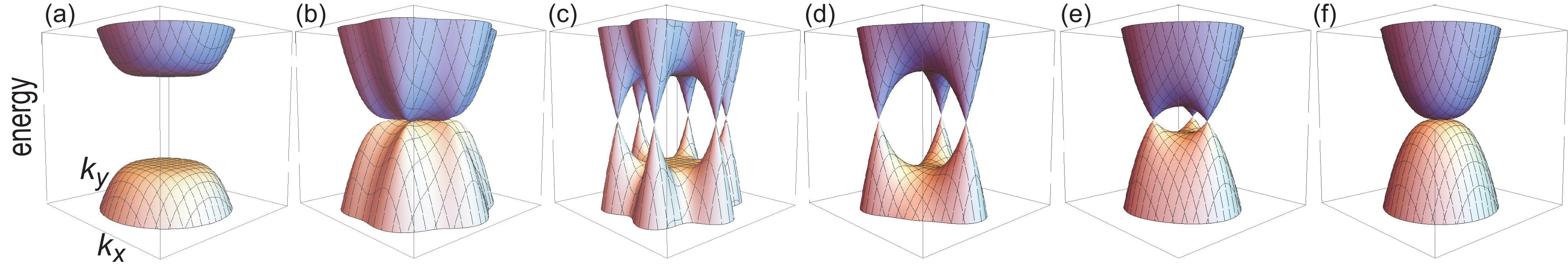}}
\caption{Bird's eye's views of the band structure on the $k_{x}k_{y}$ plane
at $k_{z}=0$ in the case of $N=3$, when the magnetic field is $\mathbf{h}=0$
for (a), $\mathbf{h}=(m,0,0)$ for (b), and $\mathbf{h}=(2m,0,0)$ for (c).
Three pairs of Weyl and anti-Weyl points are observed in (c). Similar views
at $k_{z}=(\protect\sqrt{h^{2}-m^{2}}/a_{z})\cos \protect\theta $, when $%
\mathbf{h}=2m(\cos \protect\theta ,0,\sin \protect\theta )$ with $\protect%
\theta =\frac{1}{4}\protect\pi $ for (d), $\protect\theta =\frac{4}{9}%
\protect\pi $ for (e), and $\protect\theta =\frac{1}{2}\protect\pi $ for
(f). Three Weyl points are observed in (d) and (e), which are merged into
one multiple-Weyl point in (f).}
\label{FigWeyl}
\end{figure*}

\textit{Magnetic field:} When $\mathbf{h}=\mathbf{0}$, the Hamiltonian
describes an insulator with massive Dirac electrons whose dispersion is $N$%
-th power along the $k_{x}$ and $k_{y}$ directions and linear along the $%
k_{z}$ direction: See Fig.\ref{FigWeyl}(a). Without loss of generality we
can choose $\mathbf{h}=h(\sin \theta ,0,\cos \theta )$ and $h>0$. The
Hamiltonian (\ref{BasicHamil}) yields the energy spectrum $E_{\chi \eta
}=\chi \sqrt{F+2\eta h\sqrt{G}}$ with $\chi =\pm 1$, $\eta =\pm 1$, and%
\begin{eqnarray}
F &=&(ak)^{2N}+h^{2}+m^{2}+a_{z}^{2}k_{z}^{2}, \\
G &=&m^{2}+\left( (ak)^{N}\cos N\phi \sin \theta +a_{z}k_{z}\cos \theta
\right) ^{2},
\end{eqnarray}%
where we have set $k_{x}=k\cos \phi $ and $k_{y}=k\sin \phi $. The gap is
given by $\Delta =\sqrt{m^{2}-h^{2}}$ at $\mathbf{k}=\mathbf{0}$ for $h<m$.
As $h$ increases the gap decreases and closes at the critical value $h=m$,
as shown in Fig.\ref{FigWeyl}(b). Then, for $h>m$, the band splits into $N$
Weyl points and $N$ anti-Weyl points with the linear dispersion, as shown in
Fig.\ref{FigWeyl} (c). The zero-energy solutions at these $2N$ points are
given by 
\begin{eqnarray}
ak_{x} &=&\left( h^{2}-m^{2}\right) ^{1/2N}\cos (j\pi /N)\sin \theta ,\quad
\\
ak_{y} &=&\left( h^{2}-m^{2}\right) ^{1/2N}\sin (j\pi /N)\sin \theta ,\quad
\\
a_{z}k_{z} &=&\pm \sqrt{h^{2}-m^{2}}\cos \theta ,  \label{EqKz}
\end{eqnarray}%
with $j=1,\cdots ,2N$.

\begin{figure*}[t]
\centerline{\includegraphics[width=0.99\textwidth]{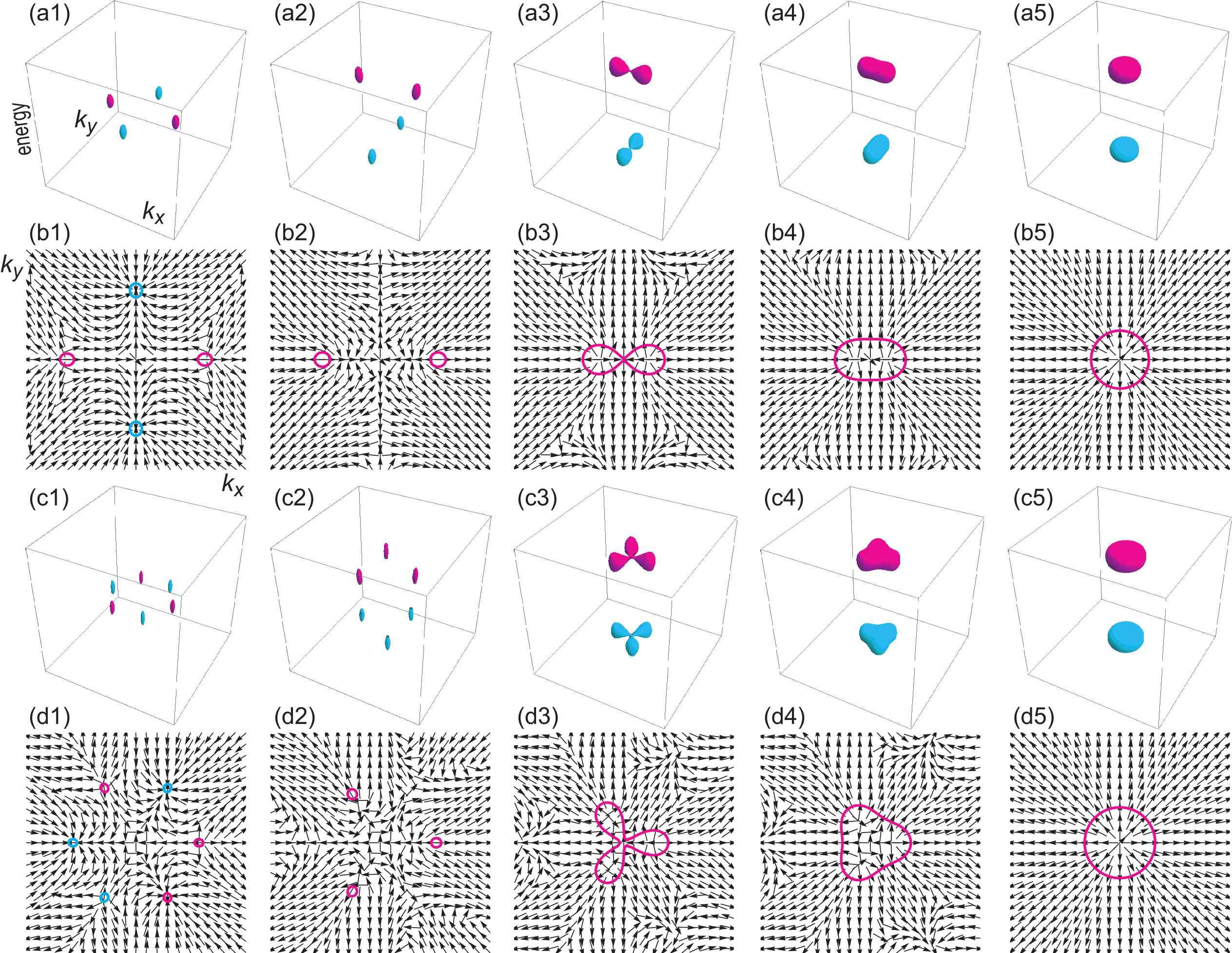}}
\caption{Bird's eye's views of the almost zero-energy surfaces ($E=\protect%
\delta $) of the Hamiltonian in the case of $N=2$ for $\mathbf{h}=2m(\cos 
\protect\theta ,0,\sin \protect\theta )$ with (a1) $\protect\theta =0$, (a2) 
$\protect\theta =\frac{1}{4}\protect\pi $, (a3) $\protect\theta =\frac{4}{9}%
\protect\pi $, (a4) $\protect\theta =\frac{17}{36}\protect\pi $, (a5) $%
\protect\theta =\frac{1}{2}\protect\pi $. (b1)-(b5) Normalized Berry
curvature $(\Omega _{x},\Omega _{y})/\protect\sqrt{\Omega _{x}^{2}+\Omega
_{y}^{2}}$ and the almost zero-energy surfaces\ corresponding to (a1)-(a5).
Two pairs of monopoles (in magenta) and antimonopoles (in cyan) are observed
in (b1), while two monopoles are observed in (b2) and (b3). They approach
one another very much in (b4), and are merged into one monopole in (b5).
Magenta circles become larger as $\protect\theta $ approaches $\frac{1}{2}%
\protect\pi $, indicating that the dispersion becomes flatter. (c1)-(c5)
Bird's eye's view of the almost zero-energy surface of the Hamiltonian in
the case of $N=3$. Magnetic field is the same as that of (a1)-(a5).
(d1)-(d5) Normalized Berry curvature corresponding to (c1)-(c5).}
\label{FigBerry}
\end{figure*}

The Fermi surface is composed of the zero-energy points. All of them are on
a single plane parallel to the $k_{x}k_{y}$ plane at $k_{z}$ fixed by (\ref%
{EqKz}). We show the almost zero-energy surface given by $E=\delta $ for a
fixed small value of $\delta $ at $h=2m$ in Fig.\ref{FigBerry}(a1) and (c1)
for $N=2$ and $3$, respectively. Each of these points on the plane at $%
k_{z}>0$ ($k_{z}<0$) is a monopole (antimonopole) with the monopole charge $%
\pm 1$.

This can be seen in the standard way. Namely, we solve for the eigenstate $%
\left\vert \psi \right\rangle $ in the Hamiltonian (\ref{BasicHamil}),
calculate the Berry connection by $A_{i}\left( \mathbf{k}\right) =-i\langle
\psi |\partial _{i}\left\vert \psi \right\rangle $ and the Berry curvature
by $\mathbf{\Omega }\left( \mathbf{k}\right) =\nabla \times \mathbf{A}(%
\mathbf{\mathbf{k}})$, where $\partial _{i}=\partial /\partial k_{i}$. We
show the Berry curvature at $h=2m$ in Fig.\ref{FigBerry}(b1) and (d1) for $%
N=2$ and $3$, respectively. Each Weyl (anti-Weyl) point has a hedgehog
(anti-hedgehog) structure and possess the unit (minus unit) of monopole
charge in the Berry curvature. The $N$ Weyl (anti-Weyl) points merge into
one Weyl (anti-Weyl) point at $\mathbf{h}=(0,0,h)$. The coordinate of the
point is given by $(0,0,\pm \sqrt{h^{2}-m^{2}})$, which we call the north or
south pole. We clearly see how the $N$ hedgehog structures merge into one
hedgehog structure in Fig.\ref{FigBerry}. The dispersion along the $k_{x}$
and $k_{y}$ axes is the $N$-th power, which manifests that the Fermi surface
is largely flattened along the $k_{x}$ and $k_{y}$ directions compared with
the $k_{z}$ direction as shown in Fig.\ref{FigBerry}(a5) and (c5).

\textit{Anomalous Hall effects:} It is known that a pair of Weyl points
contributes to the anomalous Hall conductivity, which is proportional to the
distance of the pair\cite{Hosur}. It is due to the fact that the system has
a nontrivial Chern number as a function of $k_{z}$. We seek whether similar
phenomena exist in the present system.

The anomalous Hall conductance is given by $\sigma _{ij}=(e^{2}/\hbar
)\varepsilon _{ij\ell }\int C\left( k_{\ell }\right) dk_{\ell }$, where $%
C\left( k_{\ell }\right) $ is the Chern number at $k_{\ell }$ and defined by 
$C\left( k_{\ell }\right) =\sum_{\nu }C_{\nu }\left( k_{\ell }\right) $ with 
$C_{\nu }\left( k_{\ell }\right) =(1/2\pi )\varepsilon _{\ell ij}\int
dk_{i}dk_{j}\Omega _{\ell }\left( \mathbf{k}\right) $. The summation $%
\sum_{\nu }$ runs over the occupied bands indexed by $\nu $. Here, $C_{\nu
}\left( k_{\ell }\right) $ is the total Berry magnetic flux through the $%
k_{i}k_{j}$ plane that comes from the band indexed by $\nu $.

A monopole charge is given by the surface integral of the Berry magnetic
flux surrounding a Weyl point. We take a cylindrical surface containing a
single Weyl point whose thickness is infinitesimally small and whose radius
is infinitely large. Let the vector parallel to the cylindrical axis be $%
(\sin \vartheta ,0,\cos \vartheta )$. We focus on the Chern number $%
C(k_{\vartheta })=C(k_{x})\sin \vartheta +C(k_{z})\cos \vartheta $ with $%
k_{\vartheta }=k_{x}\sin \vartheta +k_{z}\cos \vartheta $. The contribution
from the side of the cylinder is zero. The contribution from the top and
bottom surfaces are the Chern numbers $C\left( k_{\vartheta }+\delta \right) 
$ and $C\left( k_{\vartheta }-\delta \right) $, respectively, and hence the
total contribution is $C\left( k_{\vartheta }+\delta \right) -C\left(
k_{\vartheta }-\delta \right) =\pm 1$ for a monopole or an antimonopole.
Since there are no monopoles and no antimonopoles as $k_{\omega }\rightarrow
-\infty $, we set $\lim_{k_{\omega }\rightarrow -\infty }C\left(
k_{\vartheta }\right) =0$. In this way $C\left( k_{\vartheta }\right) $ is
obtained as a function of $k_{\vartheta }$.

The Chern number $C\left( k_{z}\right) $ is calculated as follows. For the
the band with $\chi =-1$ and $\eta =-1$, we obtain explicitly as 
\begin{equation}
C_{\nu }\left( k_{z}\right) =\left\{ 
\begin{array}{ccc}
N/2 & \text{for} & a_{z}|k_{z}|>\sqrt{h^{2}-m^{2}}\cos \theta \\ 
-N/2 & \text{for} & a_{z}|k_{z}|<\sqrt{h^{2}-m^{2}}\cos \theta%
\end{array}%
\right. .
\end{equation}%
It changes the sign at the position of the Weyl and anti-Weyl points. On the
other hand, the Chern number does not change as a function of $k_{z}$ for
the band with $\chi =-1$ and $\eta =1$, $C_{\nu}\left( k_{z}\right) =-N/2$.
The total Chern number is given by the sum of these two contributions and
given by 
\begin{equation}
C\left( k_{z}\right) =\left\{ 
\begin{array}{ccc}
0 & \text{for} & a_{z}|k_{z}|>\sqrt{h^{2}-m^{2}}\cos \theta \\ 
-N & \text{for} & a_{z}|k_{z}|<\sqrt{h^{2}-m^{2}}\cos \theta%
\end{array}%
\right. .
\end{equation}%
Similarly we can calculate $C\left( k_{x}\right) $ and $C\left( k_{y}\right) 
$.

It follows that the anomalous Hall conductance is given as%
\begin{equation}
\sigma _{xy}=N\frac{e^{2}}{\pi h}\sqrt{h^{2}-m^{2}}\cos \theta .
\end{equation}%
It is enhanced by $N$ times compared with that in normal Weyl semimetals.

\textit{Fermi arcs:} In what follows we study the case $\mathbf{h}=(0,0,h)$. 
$N$ Weyl points merge into a single multiple-Weyl point at the north or
south pole for $h>m$. To such a case we apply the same argument and find
that the monopole charge at the north (south pole) is $\pm N$. The system is
a quantum anomalous Hall insulator for each $k_{z}$ when $a_{z}|k_{z}|<\sqrt{%
h^{2}-m^{2}}$. The $N$ chiral edge modes appear at the sample edges based on
the bulk-edge correspondence. Each chiral edge must
cross the Fermi energy at a certain momentum\cite{Hosur}. These zero-energy
states are present continuously between the north and south poles, implying
the emergence of $N$ Fermi arcs connecting these poles.

\textit{Multiple-Weyl fermions:} In order to make a further study of the
monopole, we derive the effective Hamiltonian for multiple-Weyl fermions
under the magnetic field $\mathbf{h}=(0,0,h)$ with $h>m$. We first construct
a unitary transformation $U$ that diagonalizes the Hamiltonian $H\left(
0,0,k_{z}\right) $. We then transform the Hamiltonian $H\left(
k_{x},k_{y},k_{z}\right) $ by the same unitary transformation $U$. We may
extract the $2\times 2$ matrix whose eigenvalues vanish at the north and
south poles. Expanding it around $k_{z}=\pm \sqrt{h^{2}-m^{2}}$, we obtain%
\begin{eqnarray}
H_{\text{eff}} &=&-a^{N}\left( k_{+}^{N}\sigma _{-}+k_{-}^{N}\sigma
_{+}\right)  \notag \\
&&\mp \sqrt{1-\frac{m^{2}}{h^{2}}}\left( a_{z}k_{z}\mp \sqrt{h^{2}-m^{2}}%
\right) \sigma _{z}.  \label{EffecHamil}
\end{eqnarray}%
This is the effective Hamiltonian valid around the north or south pole.

We prove that the north pole has the monopole charge $N$. With the use of
the eigenstate of the Hamiltonian (\ref{EffecHamil}) we may derive the Berry
curvature as%
\begin{equation}
\left( \Omega _{x},\Omega _{y},\Omega _{z}\right) =\frac{Na^{2N}a_{z}\left(
k_{x}^{2}+k_{y}^{2}\right) ^{N-1}\left( k_{x},k_{y},Nk_{z}\right) }{2\left(
a^{N}\left( k_{x}^{2}+k_{y}^{2}\right) ^{N}+a_{z}^{2}k_{z}^{2}\right) ^{3/2}}%
.
\end{equation}%
This describes the momentum-space monopole at the north pole. It is easy to
show $\partial _{i}\Omega _{i}(\mathbf{k})=0$ except for $\mathbf{k}=0$.
Hence, we find $\partial _{i}\Omega _{i}=\rho \delta (\mathbf{k})$ with a
constant $\rho $. We make a change of the variables, $k_{x}=\left( K\sin
\theta \right) ^{1/N}\cos \phi $, $k_{y}=\left( K\sin \theta \right)
^{1/N}\sin \phi $ and $k_{z}=K\cos \theta $, with $(K,\theta ,\phi )$ being
the polar coordinate. To determine the constant $\rho $, we use the Gauss
theorem by choosing a sphere with radius $K$. We then have%
\begin{equation}
\rho =\epsilon _{ij\ell }\oint \Omega _{i}dk_{j}dk_{\ell }=\frac{N}{2}\frac{%
1}{2\pi }\oint \sin \theta d\theta d\phi =N
\end{equation}%
after a straightforward calculation.

\begin{figure}[t]
\centerline{\includegraphics[width=0.49\textwidth]{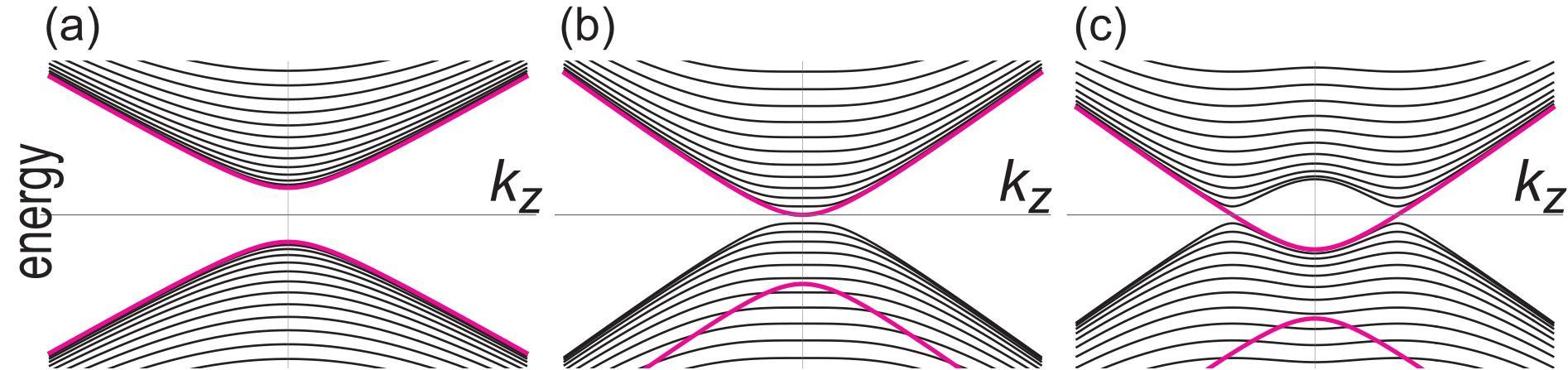}}
\caption{Landau levels as a function of $k_{z}$. (a) $h=0$ (below the critical value), where the edge spectrum is embedded in the bulk spectrum. (b)  $h=m$ (the critical
value), where a part of the edge spectrum  goes away from the Fermi energy toward the bottom. (c)  $h=2m$ (above the critical value), where the chiral edge modes appear at the Weyl and the anti-Weyl points. The bulk spectrum is colored in black, while the
edge spectrum is colored in magenta. The edge spectrum is $N$-fold degenerated.}
\label{FigLL}
\end{figure}

\textit{Landau levels:} When the magnetic field is strong enough, the
cyclotron motion in the $xy$ plane forms Landau levels\cite{Xli,Bitan,Multi}%
. As a result, the system becomes a one-dimensional system along the $k_{z}$
axis. By making the minimal substitution, the Hamiltonian is given by 
\begin{equation}
\hat{H}=\tau _{z}\left( \hbar \omega _{\text{c}}\left( \hat{a}^{\dagger
}\right) ^{^{N}}\sigma _{-}+\hbar \omega _{\text{c}}\hat{a}^{N}\sigma
_{+}+a_{z}k_{z}\sigma _{z}\right) +m\tau _{x},
\end{equation}%
where $\hat{a}$ is the Landau-level ladder operator with $[\hat{a},\hat{a}%
^{\dag }]=1$, and $\hbar \omega _{\text{c}}$ is the cyclotron energy. The
bulk spectrum is obtained, 
\begin{equation}
E_{n}^{\chi \eta }=\chi \sqrt{\frac{\left( n+N\right) !}{n!}\hbar \omega _{%
\text{c}}+\left( \sqrt{a_{z}^{2}k_{z}^{2}+m^{2}}+\eta h\right) ^{2}},
\end{equation}%
with the eigenstates being 
\begin{equation}
\psi =\left( u_{n}\left\vert n\right\rangle ,u_{n+N}\left\vert
n+N\right\rangle ,u_{n}\left\vert n\right\rangle ,u_{n+N}\left\vert
n+N\right\rangle \right) ^{t}
\end{equation}%
for $n=0,1,2,\cdots $. In addition we have $N$-fold degenerated Landau levels%
\begin{equation}
E=-h\pm \sqrt{a_{z}^{2}k_{z}^{2}+m^{2}},
\end{equation}%
with the eigenstates being $\psi =\left( 0,u_{\nu }\left\vert \nu
\right\rangle ,0,u_{\nu }\left\vert \nu \right\rangle \right) ^{t}$, where $%
\nu =0,\cdots ,N-1$. We show the Landau levels as a function of $k_{z}$ in
Fig.\ref{FigLL}. Chiral edge modes emerge for $|h|>m$, where the degeneracy
is $N$. Hence, the chiral anomaly is enhanced by $N$ times.

\textit{Lattice Hamiltonian:} It would be interesting to construct a lattice
Hamiltonian, from which the continuum Hamiltonian (\ref{BasicHamil}) follows
as a low energy theory. A simplest realization would be 
\begin{eqnarray}
H &=&-2^{N-1}t\tau _{z}\left[ \sigma _{x}\prod\limits_{j=1}^{N}\sin \left( 
\mathbf{d}_{j}^{1}\cdot \mathbf{k}\right) +\sigma
_{y}\prod\limits_{j=1}^{N}\sin \left( \mathbf{d}_{j}^{2}\cdot \mathbf{k}%
\right) \right]   \notag \\
&&+t_{z}\sigma _{z}\sin a_{z}k_{z}+m\tau _{x}+\mathbf{h}\cdot \mathbf{\sigma 
},  \label{Lattice}
\end{eqnarray}%
with $\mathbf{d}_{j}^{1}=a\left( \sin \left[ \left( 2j+1\right) \pi /2N%
\right] ,\cos \left[ \left( 2j+1\right) \pi /2N\right] ,0\right) $ and $%
\mathbf{d}_{j}^{2}=a\left( \sin \left[ j\pi /N\right] ,\cos \left[ j\pi /N%
\right] ,0\right) $, where $t$, $t_{z}$ are transfer energies, and $a$, $%
a_{z}$ are lattice constants. The energy spectrum has the zero-energy points
at the $\Gamma $ point ($0,0,0$). In the vicinity of the $\Gamma $ point,
the lattice Hamiltonian (\ref{Lattice}) is expanded and yields the continuum
Hamiltonian (\ref{BasicHamil}).

\textit{Discussions:} We have studied a merging process of $N$ monopoles
into one monopole carrying monopole charge $N$ based on the continuum
Hamiltonian. We have constructed a lattice Hamiltonian from which the
continuum Hamiltonian is derived. In the lattice Hamiltonian (\ref{Lattice}%
), $N$ represents the number of the nearest adjacent sites. Hence, the case
for $N=2$ can be embedded into the layered square lattice, while the case
for $N=3$ can be embedded into the layered triangular lattice\cite%
{CFang,BJYang}. On the other hand, it is impossible to embed the cases for $%
N\geq 4$ into the realistic lattice as far as we employ this lattice
Hamiltonian.

Finally we would argue a possible experimental study of the merging process.
When the mass parameter $m$ is large the magnetic field becomes too large to
carry out any experiments. However, our theory is valid however small $m$
may be. Recent study shows that a cubic-Dirac semimetal is realized in
quasi-one-dimensional transition-metal monochalcogenides by first-principles
calculations\cite{QLiu}, where the mass parameter is almost zero. It is easy
to realize the topological phase transition for small $m$ since the
transition occurs at $h=m$. In this sense, quasi-one-dimensional
transition-metal monochalcogenides will be an ideal playground to verify the
results in our predictions.

The author is very much grateful to N. Nagaosa for many helpful discussions
on the subject. This work is supported by the Grants-in-Aid for Scientific
Research from MEXT KAKENHI (Grant Nos.JP17K05490 and 15H05854). This work
was also supported by CREST, JST (Grant No. JPMJCR16F1).


\begin{thebibliography}{99}
\bibitem{Hasan} M. Z. Hasan and C. L. Kane, Rev. Mod. Phys. 82, 3045 (2010)

\bibitem{Qi} X.-L. Qi and S.-C. Zhang, Rev. Mod. Phys. 83, 1057 (2011)

\bibitem{Hosur} P. Hosur, X.L. Qi, C. R. Physique 14, 857 (2013).

\bibitem{JPSJReview} M. Ezawa, J. Phys. Soc. Jpn. 84, 121003 (2015).

\bibitem{Murakami} S. Murakami, New J. Phys. 9, 356 (2007).

\bibitem{Burkov} A. A. Burkov and Leon Balents, Phys. Rev. Lett. 107, 127205
(2011).

\bibitem{KK} K.-Y. Yang, Y.-M. Lu, and Y. Ran, Phys. Rev. B 84, 075129
(2011).

\bibitem{ABJ} H. B. Nielsen and M. Ninomiya, Phys. Lett. B \textbf{130}, 389
(1983).

\bibitem{Zyuzin} A. A. Zyuzin and A. A. Burkov, Phys. Rev. B 86, 115133
(2012).

\bibitem{Huang} X. Huang, L. Zhao, Y. Long, P. Wang, D. Chen, Z. Yang, H.
Liang, M. Xue, H. Weng, Z. Fang, X. Dai, and G. Chen, Phys. Rev. X X 5,
031023 (2015).

\bibitem{PYe} C.-X. Liu, P. Ye, and X.-L. Qi Phys. Rev. B 87, 235306 (2013)

\bibitem{CLZhang} C. L. Zhang, et. al Nat. Com. 7, 10735 (2016)

\bibitem{XWan} X. Wan, A. M. Turner, A. Vishwanath, and S. Y. Savrasov Phys.
Rev. B 83, 205101 (2011)

\bibitem{SYXu} S.Y. Xu, et.al, Science, 349, 613 (2015)

\bibitem{CFang} C. Fang, M. J. Gilbert, X. Dai, and B. A. Bernevig, Phys.
Rev. Lett. \textbf{108}, 266802 (2012).

\bibitem{BJYang} B.-J. Yang and N. Nagaosa, Nat. Commun. \textbf{5}, 4898
(2014).

\bibitem{Xli} X. Li, B. Roy and S. Das Sarma, Phys. Rev. B \textbf{94},
195144 (2016)

\bibitem{Huang2} S.-M. Huang, S.-Y. Xu, I. Belopolski, C.-C. Lee, G. Chang,
T.-R. Chang, B. Wang, N. Alidoust, G. Bian, M. Neupane, D. Sanchez, H.
Zheng, H.-T. Jeng, A. Bansil, T. Neupert, H. Lin, and M. Z. Hasan, Proc.
Natl. Acad. Sci. 113, 1180 (2016)

\bibitem{Jian} S.-K. Jian and H. Yao, Phys. Rev. B 92, 045121 (2015)

\bibitem{Ahn} S. Ahn, E. J. Mele, and H. Min, cond-mat/arXiv:1609.08566

\bibitem{QLiu} Q. Liu and A. Zunger, Phys. Rev. X, 7, 021019 (2017)


\bibitem{Bitan} B. Roy and J. D. Sau, Phys. Rev. B \textbf{92}, 125141
(2015).

\bibitem{Multi} M. Ezawa, Phys. Rev. B 95, 205201 (2017)

\end{thebibliography}
\end{document}